\documentclass[pra,aps,preprint,showpacs]{revtex4}
\usepackage{amssymb,amsfonts}
\usepackage{epsfig}
\usepackage{CJK}

\def\opone{\leavevmode\hbox{\small1\kern-3.8pt\normalsize1}}

\begin{document}
%\begin{CJK*}{GBK}{song}
\title{Dynamics of Geometric Discord and Measurement-Induced Nonlocality at Finite Temperature}
\author{Guo-Feng Zhang\footnote{Corresponding author. Tel:86 10 8231 7935}\footnote{Email:
gf1978zhang@buaa.edu.cn}} \affiliation{State Key Laboratory of Software Development Environment, Beihang University, Xueyuan Road No. 37, Beijing 100191, PR China; Department of Physics, School of Physics and Nuclear Energy Engineering, Beihang University, Xueyuan Road No. 37, Beijing 100191, PR China }
\author{Heng Fan, Ai-Ling Ji and Wu-Ming Liu} \affiliation{Beijing National Laboratory for Condensed Matter Physics, Institute of Physics, Chinese Academy of Sciences, Beijing 100190, PR China}

%%%%%%%%%%%%%%%%%%%%%%%%%%%%%%%%%%%%%%%%%%%%%%%%%%%%%%%%%%%%%%%%%%
\begin{abstract}
By using geometric measure of discord (\texttt{GMOD})[Phys. Rev. Lett, 105, 109502 (2010)] and measurement-induced nonlocality (\texttt{MIN})[Phys. Rev. Lett, 106, 120401 (2011)], we investigate quantum correlation of a pair of two-level
systems, each of which is interacting with a reservoir at finite temperature $T$. We show that, for a broad class of states of the system, \texttt{GMOD} and \texttt{MIN} can endure sudden death, and there is no asymptotic decay for \texttt{MIN} while asymptotic decay exists for \texttt{GMOD}. We also  give the dynamics of \texttt{GMOD} and \texttt{MIN} with respect to the temperature and illustrate their different characteristics.

Keywords: Geometric measure of discord (\texttt{GMOD}); Measurement-induced nonlocality (\texttt{MIN}); Finite temperatures; Sudden death; Asymptotic decay
\end{abstract}
%%%%%%%%%%%%%%%%%%%%%%%%%%%%%%%%%%%%%%%%%%%%%%%%%%%%%%%%%%%%%%%%%%

\pacs{03. 65. Yz, 03. 65. Ud, 42. 50. Lc}

\maketitle
%\end{CJK*}

\section{Introduction}

Quantum correlation arises from noncommutativity of operators representing states, observables, and measurements \cite{sll}. Quantum entanglement, which refers to the separability of the states, is very important in quantum information processing and can be realized in many kinds of physical systems which involve quantum correlation. Quantum entanglement has been investigated widely in past decades, while quantum correlation seems to have been seldom exploited before. An alternative classification for
quantum correlations, which is based on quantum measurements, has arisen in recent years and also plays an important role in quantum information theory \cite{sll,mpi,slu,nli}. In particular, quantum discord \cite{hol} is introduced to measure these quantum correlations. There exist indeed separable mixed states having nonzero discord, and the separable mixed states can be used to perform useful quantum tasks \cite{ada}. Evaluation of quantum
discord in general requires considerable numerical minimization and analytical expressions
are known only for certain classes of states. Luo evaluated analytically the quantum discord for a
large family of two-qubit states, and make a comparative study of the relationships between classical and
quantum correlations in terms of the quantum discord \cite{luo}. Dakic, Vedral, and Brukner propose a geometrical way of quantifying quantum discord \cite{dakic}, which is termed as \texttt{GMOD}. \texttt{GMOD} can be extended to any number of subsystems, though evaluating the measure of discord becomes progressively more difficult with the increasing of the number of subsystems
and that of their dimensionality. Moreover, Luo and Fu \cite{luo1} evaluate \texttt{GMOD} for an arbitrary state and obtain an explicit and tight lower bound. Very differently, measurement-induced nonlocality (\texttt{MIN}) \cite{luofu} has been proposed to interpret the maximum global effect caused by locally invariant measurements, the authors claim that \texttt{MIN} is in some sense dual to \texttt{GMOD}. Anyway, both \texttt{GMOD} and \texttt{MIN} are the measurement tool of quantum correlation.

The interaction of a quantum system with its environment causes the rapid destruction
of crucial quantum properties and drives the system to an incoherent state. It was shown by Yu and Eberly that entanglement of a bipartite system decays to zero in a finite time, which is called entanglement sudden death (ESD), while coherence vanishes exponentially with time to zero \cite{yut1,yut2}. Subsequently, ESD in different systems has been
made by various groups \cite{yut3,yut4,fla,ava}. Al-Qasimi and James \cite{aal} demonstrated that a broad class of mixed quantum states undergo ESD at finite temperatures. Recently, by using carefully engineered interactions between system and environments, experimental studies have been
carried out to demonstrate ESD, and ESD has been observed both in photons \cite{mpa} and in atomics ensembles \cite{jla}.

In this paper, motivated by the work on sudden death of entanglement at finite tem-
peratures \cite{aal}, we investigate whether \texttt{GMOD} and \texttt{MIN} endure sudden death, their different characteristics are shown. The qubits system is in finite temperature reservoirs, thus the reservoirs can cause excitation of qubits instead of the energy of the qubits being lost via the spontaneous decay to the environment. For a broad
class of mixed quantum states, which includes all of the states studied by Yu and Eberly and others in connection with this problem, we demonstrate that all states can endure sudden death of \texttt{GMQD} and \texttt{MIN} at finite temperatures, and there is no asymptotic decay for \texttt{MIN} while asymptotic decay exists for \texttt{GMQD}.

\section{Two-qubit model system}

We consider two two-level atoms $1$ and $2$ that present a two-qubit system and interact
with their local thermal reservoirs. There is no direct interaction between the atoms. The
effect of heat is included in our system. The dynamics of the density matrix $\rho$ describing the two qubits reads \cite{mik}
\begin{equation}
\frac{d\rho}{dt}=\frac{1}{2}(m+1)\gamma\Sigma^{2}_{i=1}\{[\sigma^{i}_{-},\rho\sigma^{i}_{+}]+[\sigma^{i}_{-}\rho,\sigma^{i}_{+}]\}+\frac{1}{2}m\gamma\Sigma^{2}_{i=1}\{[\sigma^{i}_{+},\rho\sigma^{i}_{-}]+[\sigma^{i}_{+}\rho,\sigma^{i}_{-}]\},
\end{equation}
where $\gamma$ is the spontaneous emission rate of the atom, and we assume that two atoms have the same value, $\sigma^{i}_{\pm}(i = 1; 2)$ are the rasing $(+)$ and lowering $(-)$ operators of atom $i$, defined as $\sigma^{i}_{+}=|1\rangle\langle0|_{i}$, $\sigma^{i}_{-}=|0\rangle\langle1|_{i}$, $m$ is the mean occupation number of the reservoir and it also is assumed to be the same for both atoms. On the right hand side of equation (1), the first term describes the depopulation of the atoms due to simulated and spontaneous emission, while the second term corresponds to the reexcitations caused by the finite temperature.

We consider the following initial state described by
\begin{equation}
\rho[0]=\left(
  \begin{array}{cccc}
    a_{0} & 0 & 0 & w_{0} \\
    0 & b_{0} & z_{0} & 0 \\
    0 & z^{*}_{0} & c_{0} & 0 \\
    w^{*}_{0} & 0 & 0 & d_{0} \\
  \end{array}
\right).
\end{equation}
One may note that the above state retains its form under the equation (1), the nonzero elements of the matrix $\rho[t]$ can be written as
\begin{eqnarray}
a[t]&=&\frac{1}{(1+2m)^{2}}\{m^{2}+mX[1+a_{0}+2ma_{0}-d_{0}(1+2m)]\nonumber\\&+&X^{2}[a_{0}+m(-1+3a_{0}+d_{0})+m^{2}(-1+2a_{0}+2d_{0})]\},\nonumber\\
b[t]&=&\frac{1}{(1+2m)^{2}}\{X[m(1+m)+(b_{0}+a_{0}(1+m)+m(-d_{0}+2b_{0}(1+m)-2c_{0}(1+m)))]\nonumber\\&+&X^{2}[-a_{0}(1+m)(1+2m)+m(1+m-d_{0}(1+2m))]\},\nonumber\\
c[t]&=&\frac{1}{(1+2m)^{2}}\{X[m(1+m)(1-a_{0})+(a_{0}(1+m)^{2}+c_{0}(1+2m+2m^{2})\nonumber\\&-&m(d_{0}+2b_{0}(1+m)))]+X^{2}[-a_{0}(1+m)^{2}+m(1+m-d_{0}(1+2m))]-X^{3}ma_{0}(1+m)\},\nonumber\\
d[t]&=&\frac{1}{(1+2m)^{2}}\{X[(1+m)^{2}+(-1-m)(b_{0}+c_{0}-2d_{0}m+2a_{0}(1+m))]\nonumber\\&+&X^{2}[a_{0}+(-1+3a_{0}+d_{0})m+(-1+2a_{0}+2d_{0})m^{2}],\nonumber\\
w[t]&=&Xw_{0},z[t]=Xz_{0},
\end{eqnarray}
where $X=\exp[-t\gamma(1+2m)]$ and $a[0]=a_{0}$, etc.

\section{Dynamics of \texttt{GMOD} and \texttt{MIN}}
In this section, the dynamics of \texttt{GMOD} and \texttt{MIN} are considered. Taking advantage of Al-Qasimi and James procedure \cite{aal} judging whether ESD occurs, we want to investigate whether the \texttt{GMOD} and \texttt{MIN} endure sudden death.

According to Eq.(9) and Eq.(16) in Ref.\cite{dakic}, we can get the \texttt{GMOD} of $\rho[t]$
\begin{eqnarray}
\texttt{GMOD}[t]&=&\frac{1}{4}\{2(a[t]-c[t])^{2}+2(b[t]-d[t])^{2}+8(|w[t]|^{2}+|z[t]|^{2})\nonumber \\ &-&\max(2(a[t]-c[t])^{2}+2(b[t]-d[t])^{2},4(|w[t]|-|z[t]|)^{2},4(|w[t]|+|z[t]|)^{2})\},
\end{eqnarray}
based on Eq.(7) in Ref.\cite{luofu}, we obtain
\begin{eqnarray}
\texttt{MIN}[t]&=&\frac{1}{4}\{(a[t]-b[t]-c[t]+d[t])^{2}+8(|w[t]|^{2}+|z[t]|^{2})\nonumber \\ &-&4\min(\frac{1}{4}(a[t]-b[t]-c[t]+d[t])^{2},(|w[t]|-|z[t]|)^{2},(|w[t]|+|z[t]|)^{2})\},
\end{eqnarray}
Eq.(4) and Eq.(5) cannot be solved in closed form, however, we could consider the solutions of $\texttt{GMOD}[t]=0$ and $\texttt{MIN}[t]=0$ to make sure whether quantum discord and measurement-induced nonlocality sudden death occurs. As we have mentioned, at $t =0$, $X =1$, and at $t =\infty$, $X =0$. If the these two quantities decay to zero in a finite
time, the solutions of $\texttt{GMOD}[t]=0$ and $\texttt{MIN}[t]=0$ must lie in the range $0<X<1$. At $t=0$, the equations take the following value
\begin{eqnarray}
\texttt{GMOD}[0]&=&\frac{1}{4}\{2(a_{0}-c_{0})^{2}+2(b_{0}-d_{0})^{2}+8(|w_{0}|^{2}+|z_{0}|^{2})\nonumber \\ &-&\max(2(a_{0}-c_{0})^{2}+2(b_{0}-d_{0})^{2},4(|w_{0}|-|z_{0}|)^{2},4(|w_{0}|+|z_{0}|)^{2})\},
\end{eqnarray}
\begin{eqnarray}
\texttt{MIN}[0]&=&\frac{1}{4}\{(a_{0}-b_{0}-c_{0}+d_{0})^{2}+8(|w_{0}|^{2}+|z_{0}|^{2})\nonumber \\ &-&4\min(\frac{1}{4}(a_{0}-b_{0}-c_{0}+d_{0})^{2},(|w_{0}|-|z_{0}|)^{2},(|w_{0}|+|z_{0}|)^{2})\},
\end{eqnarray}
both of them are positive. At $t=\infty$, $X=0$, $\texttt{GMOD}(\infty)=0$ , while $\texttt{MIN}(\infty)=[1+ma_{0}(1+m)]^{2}/[4(1+2m)^{4}]$. Hence, there is $X=0$ solution for $\texttt{GMOD}[t]=0$, i.e., geometrical quantum discord endures asymptotical decay. The
fact that $\texttt{GMOD}[t]$ has a positive value at $X=1$, zero value at $X=0$, and $\texttt{GMOD}[t]$ is continuous,
implies that geometrical quantum discord sudden death will not occur. Similar analysis can lead to the result that no sudden death for measurement-induced nonlocality and the result that no asymptotical decay for general states unless for the states which meets $a_{0}=-1/(m+m^{2})$ (different from \texttt{GMOD}).

%We think it is the concurrence must be maximum
%between \emph{zero} and other values that make it undergo sudden death. While any one of \texttt{GMOD} and \texttt{MIN} must not like concurrence. %

\begin{figure}
\begin{center}
\epsfig{figure=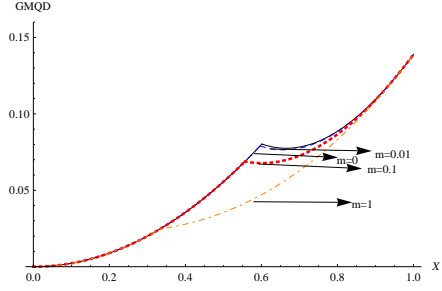,width=0.35\textwidth}
\end{center}
\caption{(Color online) Geometrical quantum discord (\texttt{GMOD}) vs $X$ for $\rho_{YE}$ when $\alpha=1/2$.}
\end{figure}
\begin{figure}
\begin{center}
\epsfig{figure=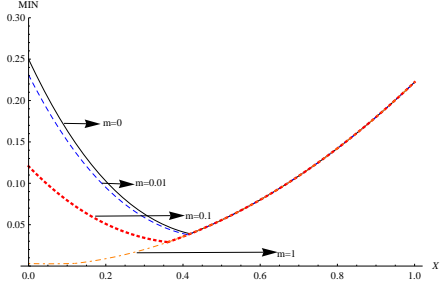,width=0.35\textwidth}
\end{center}
\caption{(Color online) Measurement-Induced Nonlocality (\texttt{MIN}) vs $X$ for $\rho_{YE}$ when $\alpha=1/2$.}
\end{figure}

Especially, we consider the states of the following form
\begin{equation}
\rho_{YE}=\frac{1}{3}\left(
  \begin{array}{cccc}
    \alpha & 0 & 0 & 0 \\
    0 & 1 & 1 & 0 \\
    0 & 1 & 1 & 0 \\
    0 & 0 & 0 & 1-\alpha \\
  \end{array}
\right).
\end{equation}
For the above states, Yu and Eberly have shown that for $0\leq\alpha\leq1/3$, the entanglement is long lived at zero temperature \cite{yut1}. In figure 1 and figure2, we give the dynamics of \texttt{GMOD} and \texttt{MIN} associated these states. The dynamics evolutions with respect to $X$ are alike except the \emph{turning point} for  different $\alpha$. We only give the result when $\alpha=1/2$. We can see that \texttt{GMOD} is the same for different $m$ (it stands for
temperature) when $X=0$ and $X=1$, while \texttt{MIN} is the same only when $X=1$.

\section{Conclusions}
In this paper, we study the evolution of geometrical quantum discord (\texttt{GMOD}) and measurement-induced nonlocality (\texttt{MIN}) in a system consisting of qubits at finite temperature reservoirs. For a class of ``X"
state, \texttt{GMOD} and \texttt{MIN} are immune to sudden death, and  even more, there is no asymptotic decay for \texttt{MIN}. We also  give the dynamics of \texttt{GMOD} and \texttt{MIN} with respect to the temperature and illustrate their different characteristics.

\section{acknowledgements}
This work was supported by the National Science Foundation of China under Grants No. 11174024, 10874013, and 10904165, as well as by the NKBRSFC under Grants No. 2010CB922904 and No. 2011CB921500,  also supported by State Key Laboratory of Software Development  Environment of BUAA Grants No. SKLSDE-2011ZX-17. Heng Fan  acknowledges the support of the National Science Foundation of China under Grant No. 10974247.
Wu-Ming Liu acknowledges the support of the National Science Foundation of China under Grant No. 10934010 and No. 60978019.

\end{document}